# Non-Boolean Probabilities and Quantum Measurement

Gerd Niestegge
Zillertalstrasse 39, D-81373 Muenchen, Germany

*Abstract*. A non-Boolean extension of the classical probability model is proposed. The non-Boolean probabilities reproduce typical quantum phenomena. The proposed model is more general and more abstract, but easier to interpret than the quantum-mechanical Hilbert space formalism and exhibits a particular phenomenon (state-independent conditional probabilities) which may provide new opportunities for an understanding of the quantum measurement process. Examples of the proposed model are provided, using Jordan operator algebras.

## 1. Introduction

The statistical interpretation is a later add-on to quantum mechanics and is neither obvious in the quantum mechanical Hilbert space formalism nor in line with Kolmogorov's measure-theoretical axioms of probability theory. With Kolmogorov's axioms, it is assumed that the events form a Boolean lattice while von Neumann[2] already pointed out that this does not hold for the quantum-mechanical events and, moreover, Kolmogorov's axioms imply the so-called Bell inequalities[1] which do not hold in quantum mechanics and were disproved by several physical experiments.

The paper presents a new axiomatic model using probabilistic interpretations from the very beginning, covering quantum mechanics (with a certain exception) as well as Kolmogorov's model and revealing a new phenomenon unknown in both these existing theories: non-trivial *state-independent conditional probabilities*. This phenomenon is called *statistical predictability*. In Kolmogorov's model, the conditional probability of an event $F$ under another event $E$ becomes independent of the underlying probability measure (state) only in the two trivial cases where either $E$ implies $F$ or $E$ implies the negation of $F$.

The *incompatibility* of events, well-known from quantum mechanics (non-commuting projections), is defined in a very basic probabilistic way. In Kolmogorov's model, all events are mutually compatible, and the new axiomatic model, indeed, reduces to Kolmogorov's model if it is assumed that all events are mutually compatible or if it is assumed that the events form a Boolean lattice. The new model can therefore be considered a non-Boolean extension of Kolmogorov's probability theory.

The wave-like interference phenomena of quantum physics are a further typical feature of non-Boolean probabilities.

The non-trivial state-independent conditional probabilities may provide opportunities for a new understanding of quantum measurement. Different aspects and views of the quantum measurement problem are presented in [5] and [17].

## 2. Orthospaces and states

With Kolmogorov's model, the events form a $\sigma$-algebra which is a special type of Boolean lattice where countably infinite sums exist. Since finite additivity is sufficient for the purpose of the present paper, an algebraic structure generalizing a Boolean lattice and rendering possible the consideration of states (finitely additive probability measures) on it is needed. A set $\mathcal{E}$ is therefore considered with distinguished elements 0 and $\mathbb{I}$, a relation $\perp$ and a partial binary operation $+$ such that for $D,E,F \in \mathcal{E}$:

(OS1)  $E \perp F \Rightarrow F \perp E$

(OS2)  $E+F$ is defined for $E \perp F$, and $E+F=F+E$

(OS3)  $D \perp E,\ D \perp F,\ E \perp F \Rightarrow D \perp E+F,\ F \perp D+E$ and $D+(E+F)=(D+E)+F$

(OS4)  $0 \perp E$ and $E+0=E$ for all $E \in \mathcal{E}$

(OS5)  For every $E \in \mathcal{E}$ there exists a unique $E' \in \mathcal{E}$ such that $E \perp E'$ and $E+E'=\mathbb{I}$

(OS6)  $E \perp F' \Leftrightarrow$ There exists a $D \in \mathcal{E}$ such that $E \perp D$ and $E+D=F$

A set $\mathcal{E}$ with the above operations satisfying (OS1-6) is called an *orthospace*. We say "$E$ and $F$ are *orthogonal*" for $E \perp F$, and $E'$ is called the *complement* of $E$. Then $0'=\mathbb{I}$, and $E''=E$ for each $E$ in an orthospace $\mathcal{E}$, which immediately follows from (OS2,4,5).

A further relation $\prec$ is defined on an orthospace $\mathcal{E}$ via: $E \prec F :\Leftrightarrow E \perp F'$ ($E,F \in \mathcal{E}$). Then $E \prec F$ if and only if $\mathcal{E}$ contains an element $D$ such that $D \perp E$ and $F=E+D$, and we have $0 \prec E \prec \mathbb{I}$ for all $E \in \mathcal{E}$. The relation $\prec$ is reflexive by (OS4) or (OS5), but is not an order relation since it is neither anti-symmetric nor transitive in general. This is a major difference to the structures usually considered as quantum logics[2,14,16] where an order relation is assumed from the very beginning. An orthospace is a rather weak structure that, without the further postulates of sec. 3 concerning the states on it, would remain meaningless.

A *state* on an orthospace $\mathcal{E}$ is a map $\mu:\mathcal{E} \to [0,1]$ such that $\mu(\mathbb{I})=1$ and $\mu(E+F)=\mu(E)+\mu(F)$ for all orthogonal pairs $E,F \in \mathcal{E}$. Then $\mu(0)=0$, and $\mu$ is additive for each finite family of pairwise orthogonal elements in $\mathcal{E}$. (OS6) ensures that $\mu(E) \le \mu(F)$ for $E \prec F$.

The elements $E \in \mathcal{E}$ are interpreted as *events* and will be called so in the following. Orthogonality means that the events exclude each other. The (only partially defined) operation $+$ is interpreted as the *or* connection of mutually exclusive events, $E'$ is the *negation* of $E$. For a state $\mu$, the interpretation of the real number $\mu(E)$ is that of the *probability* of the event $E$ in the state $\mu$.

## 3. Unique conditional probabilities

The concept of conditional probabilities well-known from Kolmogorov's model is now extended to the very general situation of orthospaces and states.

**DEFINITION 3.1:** *Let $\mu$ be a state on an orthospace $\mathcal{E}$ and $E \in \mathcal{E}$ with $\mu(E)>0$. If $\nu$ is another state such that $\nu(F)=\mu(F)/\mu(E)$ holds for all $F \in \mathcal{E}$ with $F \prec E$, then $\nu$ is called a conditional probability of $\mu$ under $E$.*

Essential shortcomings of this conditional probability are that such a state $\nu$ may not exist at all and that, if such a state exists, there may be many others. The requirement that unique conditional probabilities must exist guides us to the following definition of *UCP spaces*.

**DEFINITION 3.2:** *A UCP space is an orthospace $\mathcal{E}$ satisfying the following two axioms*:

(UC1)   *If $E,F \in \mathcal{E}$ and $E \neq F$, then there is a state $\mu$ with $\mu(E) \neq \mu(F)$.*

(UC2)   *For each state $\mu$ and $E \in \mathcal{E}$ with $\mu(E)>0$, there exists one and only one conditional probability $\mu_E$ of $\mu$ under $E$.*

$\mu_E(F)$ is the probability of the event $F$ in the state $\mu$ after the event $E$ has been observed. Using the same terminology as in mathematical probability theory, we will also write $\mu(F|E)$ for $\mu_E(F)$ in the sequel. If $\mu(E)=1$, then $\mu_E=\mu$ and $\mu(F|E)=\mu(F)$ for all $F \in \mathcal{E}$.

After having observed a sequence of $n$ events $E_1,E_2,...,E_n$ ($n>1$), the probability of a further event $F$ in the state $\mu$ is inductively defined via $\mu_{E1,E2,...,En}(F):=(\mu_{E1,E2,...,En-1})_{En}(F)$ if $\mu_{E1,E2,...,En-1}(E_n)>0$. We also write $\mu(F|E_1,E_2,...,E_n)$ for $\mu_{E1,E2,...,En}(F)$.

There is a state $\mu$ with $\mu(E)=1$ for each event $E \neq 0$ in a UCP space, since from (UC1) we get a state $\nu$ with $\nu(E)>0$, and then choose $\mu=\nu_E$.

(UC1) implies a few further algebraic properties of the orthospace $\mathcal{E}$ in addition to (OS1-6). If $E \perp F'$, there exists a $D \in \mathcal{E}$ such that $E \perp D$ and $E+D=F$, but (OS6) does not require that $D$ is uniquely determined. The uniqueness of $D$ now follows from (UC1): $E+D_1=F=E+D_2$, then $\mu(E)+\mu(D_1)=\mu(F)=\mu(E)+\mu(D_2)$ for all states $\mu$, hence $\mu(D_1)=\mu(D_2)$ and $D_1=D_2$.

Moreover, if $E \prec F$ and $F \prec E$ for $E,F \in \mathcal{E}$, then $E=F$; i.e. the relation $\prec$ is anti-symmetric: If $F=E+D_1$ and $E=F+D_2$, then $\mu(F)=\mu(E)+\mu(D_1)=\mu(F)+\mu(D_2)+\mu(D_1)$, therefore $\mu(D_1)=\mu(D_2)=0$ for all states $\mu$, and $D_1=D_2=0$ by (UC1). Note that the relation $\prec$ need not be transitive so far.

Furthermore: $E \perp E \Leftrightarrow E \perp \mathbb{1} \Leftrightarrow E=0$ (If $E \perp E$, then $E \perp E+E'=\mathbb{1}$ by (OS3,5). If $E \perp \mathbb{1}$, then $E \perp 0'$ and $E' \perp 0$, i.e. $E \prec 0$ and $0 \prec E$, hence $E=0$.). The following lemma will be needed later.

**LEMMA 3.3:** *Let $\mathcal{E}$ be an orthospace such that the only elements which are orthogonal to any $D \neq 0$ are $0$ and $D'$. Then either $\mathcal{E}$ has one of the two shapes $\{0, \mathbb{1}\}$ and $\{0, \mathbb{1}, E, E'\}$, or $\mathcal{E}$ is not a UCP space.*

P r o o f : We assume that $\mathcal{E}$ is a UCP space containing an element $E$ with $0 \neq E \neq \mathbb{1}$ and at least one element $F$ different from $0, \mathbb{1}, E, E'$. Then there is a state $\mu$ with $\mu(F)>0$. If $\mu_F$ is the conditional probability of $\mu$ under $F$,

$$\nu(D):=\begin{cases} \mu_F(D) & \text{for } D \neq E, E' \\ 1-\mu_F(D) & \text{for } D = E, E' \end{cases}$$

defines a state $\nu$, since we have to observe only $\nu(D)+\nu(D')=1$ for all events $D$. Thus, $\nu$ is a conditional probability of $\mu$ under $F$ and differs from $\mu_F$ unless $\mu_F(E)=1/2$; in this case define $\nu(E)=1$ and $\nu(E')=0$.

## 4. Statistical predictability

An *atom* in an orthospace $\mathcal{E}$ is an element $0 \neq D \in \mathcal{E}$ such that $\mathcal{E} \ni F \prec D$ implies that either $F=0$ or $F=D$. If $D$ is an atom in the UCP space $\mathcal{E}$, then there is a unique state $\rho_D$ such that $\rho_D(D)=1$; $\rho_D$ is the conditional probability of any $\mu$ with $\mu(D)>0$ under $D$, which does not depend on $\mu$ in this case:

$$\rho_D(F)=\mu_D(F)=\mu(F|D) \text{ for all states } \mu \text{ with } \mu(D)>0.$$

Here, we encounter a very interesting phenomenon: a conditional probability that does not depend on the state. This phenomenon is named *statistical predictability* in the next definition.

**DEFINITION 4.1:** *An event $E$ in a UCP space $\mathcal{E}$ is called statistically predictable under the event sequence $0 \neq F_1,...,F_n \in \mathcal{E}$ ($n \geq 1$) if $\mu_1\big(E \big| F_1,...,F_n\big) = \mu_2\big(E \big| F_1,...,F_n\big)$ for all states $\mu_1$, $\mu_2$ with $\mu_j\big(F_1\big) \neq 0 \neq \mu_j\big(F_{k+1} \big| F_1,...,F_k\big)$ for $k=1,...,n-1$ ($j=1,2$). In this case, the state-independent conditional probability of $E$ under $F_1,...,F_n$ is denoted by $I\!\!P\big(E \big| F_1,...,F_n\big)$.*

Because of its state-independence, the probability $I\!\!P(E|F_1,...,F_n)$ is completely determined by the algebraic structure of the orthospace $\mathcal{E}$ and thus seems to be less a stochastic than a logical phenomenon. With classical probabilities, the state (which is the probability measure in this case) represents a lack of knowledge about the system under consideration, and with complete knowledge all probabilities would reduce to 1 and 0. This is not possible with $I\!\!P(E|F_1,...,F_n)$ since it is state-independent already.

If $E$ is statistically predictable under $F_n$, then $E$ is statistically predictable under the event sequence $F_1,...,F_n$ with $I\!\!P(E|F_1,...,F_n) = I\!\!P(E|F_n)$, i.e. any previous observations $F_1,...,F_{n-1}$ can be ignored in this case and particularly if $F_n$ is an atom (Use the inductive definition of the conditional probability of $E$ under $F_1,...,F_n$ in a state $\mu$).

Moreover, $I\!\!P(F_n|F_1,...,F_n)=1$ always holds, but $I\!\!P(F_k|F_1,...,F_n)$ need not equal 1 for $k<n$ (e.g. if $F_k$ is statistically predictable under $F_n$ with $I\!\!P(F_k|F_n)\neq1$; see sec. 7). This means that, if the same property is observed a second time directly after the first time, the second observation will always provide the same result as the first one. However, if the second observation is not repeated directly after the first one, i.e. if another property has been tested in between, there is a chance that the second observation provides another result than the first one. The information gained from the first observation seems to have been destroyed by testing the other property.

If $F_k$ is an atom (for a $k$ with $1 \leq k \leq n$), we get:

$$I\!\!P\big(E \big| F_1,...,F_n\big) = \rho_{F_k}\big(E | F_{k+1},...,F_n\big).$$

This means that, after the observation of an event which is an atom, all previous observations become meaningless for predictions concerning future observations. Nevertheless, these predictions do not become deterministic as they would after the observation of an atom in Kolmogorov's classical model (in this case, an atom is a set containing one single point).

A state that provides only the probabilities 0 and 1 ($\mu(E) \in \{0,1\}$ for all events $E$) is called *dispersion-free*.

**LEMMA 4.2:** *If a UCP space $\mathcal{E}$ contains two elements $E,F$ such that $E$ is statistically predictable under $F$ with $0 < I\!\!P(E|F) < 1$, then $\mu(F)=0$ for all dispersion-free states $\mu$ on $\mathcal{E}$.*

Proof: We assume $\mu(F)=1$. Then $\mu(E) = \mu(E|F) = I\!\!P(E|F)$ and $\mu$ is not dispersion-free.

LEMMA 4.2 excludes that the non-trivial cases of the statistical predictability can be modeled by so-called *hidden variable theories*. Despite of its simplicity, it can be considered a very general version of a Kochen-Specker[11] type theorem valid not only for the quantum mechanical standard model but for all UCP spaces. We will later see how the quantum mechanical standard model fits into the UCP space model and that non-trivial cases of the statistical predictability do exist.

## 5. Compatibility and Boolean lattices

Every Boolean lattice (= Boolean algebra) becomes a UCP space by defining: $E \perp F :\Leftrightarrow E \wedge F = \varnothing$, and $E + F := E \vee F$ for $E \perp F$. Then

$$\mu(E|F) = \frac{\mu(E \wedge F)}{\mu(F)}$$

for a state $\mu$ and an event $F$ with $\mu(F) > 0$. This coincides with the classical definition of a conditional probability. The following classical formula then holds for $0 < \mu(F) < 1$:

$$\mu(E) = \mu(E|F)\mu(F) + \mu(E|F')\mu(F') \,.$$

An event $E$ is statistically predictable under an event $F$ if and only if either $F \leq E$ or $F \leq E'$. $I\!P(E|F) = 1$ if $F \leq E$, and $I\!P(E|F) = 0$ if $F \leq E'$. This means that non-trivial cases of the statistical predictability do not occur within the Boolean lattices. The above classical formula motivates the following definition of *compatibility*.

**DEFINITION 5.1:** *A pair of events $E,F$ in a UCP space $\mathcal{E}$ is called compatible if*

$$\mu(E) = \mu(E|F)\mu(F) + \mu(E|F')\mu(F') \text{ for all states } \mu \text{ with } 0 < \mu(F) < 1, \text{ and}$$
$$\mu(F) = \mu(F|E)\mu(E) + \mu(F|E')\mu(E') \text{ for all states } \mu \text{ with } 0 < \mu(E) < 1.$$

Obviously, the events $E$ and $F$ are compatible if and only if $E$ and $F'$ are compatible. Moreover, $E$ and $F$ are compatible if one of the relations $E \perp F$ or $E \prec F$ holds.

DEFINITION 5.1 provides a purely probabilistic concept of compatibility. Before studying its relation to the well-known quantum mechanical concept of compatibility, we shall first show that it characterizes the Boolean cases among the UCP spaces in a certain sense. Of course, in a Boolean lattice, all events are mutually compatible.

**THEOREM 5.2:** *If all events in a UCP space $\mathcal{E}$ are mutually compatible, then there is an injective homomorphism of the orthospace $\mathcal{E}$ in a Boolean lattice.*

Proof: Let $\mathcal{B}$ be the space of all real-valued finitely additive functions on the UCP space $\mathcal{E}$. The set $\Omega$ that consists of the states on $\mathcal{E}$ is a convex and compact set in $\mathcal{B}$ equipped with the product topology. From the Krein-Milman theorem we then get:

$$\Omega = \overline{conv} \, ext(\Omega) \,,$$

i.e. $\Omega$ is the closed convex hull of the set of its extreme points $ext(\Omega)$.

We now show that the extreme states are dispersion-free. Assume that $\mu \in ext(\Omega)$ with $0 < \mu(E) < 1$ for some event $E$. Then due to the compatibility

$$\mu(F) = \mu(E)\mu(F|E) + \big(1 - \mu(E)\big)\mu(F|E') \text{ for all events } F,$$

which contradicts $\mu \in ext(\Omega)$.

If $E$ and $F$ are events such that $\mu(E)=\mu(F)$ for all $\mu \in ext(\Omega)$, then $\mu(E)=\mu(F)$ for all $\mu \in \Omega$ and $E=F$ by (UC1). Now let $\Gamma$ be the set of all dispersion-free states and represent an event $E$ as the set $\{\mu \in \Gamma | \mu(E)=1\}$ in the Boolean lattice formed by the power set of $\Gamma$.

In a Boolean lattice, we have

$$\mu\left(E \middle| F_1,...,F_n\right) = \frac{\mu\left(E \wedge F_1 \wedge ... \wedge F_n\right)}{\mu\left(F_1 \wedge ... \wedge F_n\right)} = \mu\left(E | F_1 \wedge ... \wedge F_n\right),$$

which does not depend on the order of the events $F_1,...,F_n$. Any permutation of them provides the same conditional probability, and it does not play any role which one among them is observed first and which one later. This does not hold in the general case. For instance, consider two atoms $F_1,F_2$ with $\mu(F_1),\mu(F_2), I\!P(F_1|F_2), I\!P(F_2|F_1) \neq 0$. Then $\mu(E|F_1,F_2)= I\!P(E|F_2)$ and $\mu(E|F_2,F_1)= I\!P(E|F_1)$. We shall later see how examples with $I\!P(E|F_2) \neq I\!P(E|F_1)$ can be found.

In Boolean lattices, the observation of the event series $F_1,...,F_n$ is identical with the observation of the single event $F_1 \wedge ... \wedge F_n$. However, in the general case, the logical "and"-operation $\wedge$ is not available for incompatible events and observing an event $F_1$ first and an event $F_2$ second becomes different from observing $F_2$ first and $F_1$ second. Timely order seems to have more significance then than in the Boolean case.

## 6. Jordan operator algebras

We shall now study a class of UCP spaces that includes the quantum-mechanical model, but also provides examples of UCP spaces that are covered neither by the quantum-mechanical nor by Kolmogorov's model. A UCP space of this class consists of the idempotent elements (projections) in a Jordan algebra.

A *Jordan algebra*[10] is a linear space $\mathcal{A}$ equipped with a (non-associative) product $\circ$ satisfying

$$X \circ Y = Y \circ X \ \text{ and } \ X \circ \left(Y \circ X^2\right) = \left(X \circ Y\right) \circ X^2$$

for all $X,Y \in \mathcal{A}$. In the present paper, only Jordan algebras over the field of real numbers are considered. As usual, $\{X,Y,Z\}$ denotes the triple product

$$\{X,Y,Z\} := X \circ \left(Y \circ Z\right) - Y \circ \left(Z \circ X\right) + Z \circ \left(X \circ Y\right)$$

of the three elements $X,Y,Z$ in $\mathcal{A}$. The following identities hold for elements $X,Y,Z$ in any real Jordan algebra:

$$\left\{\{X,Y,X\},Z,\{X,Y,X\}\right\} = \left\{X,\{Y,\{X,Z,X\},Y\},X\right\}$$

and

$$\{X,Y,X\}^2 = \left\{X,\{Y,X^2,Y\},X\right\}.$$

An element $E \in \mathcal{A}$ with $E^2=E$ is called *idempotent*. If $\mathcal{A}$ contains a (multiplicative) unit $1\!\!1$, the system $\mathcal{J}(\mathcal{A})$ of all idempotent elements in $\mathcal{A}$ forms an orthospace with $E':= 1\!\!1 - E$ and $E \perp F :\Leftrightarrow E \circ F = 0$.

A *JB algebra*[10] is a real Jordan algebra $\mathcal{A}$ that is a Banach space with a norm satisfying

$$\|X \circ Y\| \leq \|X\| \|Y\|, \ \|X^2\| = \|X\|^2 \ \text{ and } \ \|X^2\| \leq \|X^2 + Y^2\|$$

for all $X,Y \in \mathcal{A}$. The subset $\mathcal{A}_+ := \left\{ X^2 \, \middle| \, X \in \mathcal{A} \right\}$ of a JB algebra $\mathcal{A}$ is a closed convex cone, and a partial ordering is defined via: $X \leq Y \Leftrightarrow Y - X \in \mathcal{A}_+$.

For idempotent elements $E$ and $F$, $E \leq F$ is equivalent to $E \circ F = E$. With $E \perp F :\Leftrightarrow E \circ F = 0$, the relation $\prec$ then coincides with $\leq$ and becomes a partial ordering on $\mathcal{J}(\mathcal{A})$.

Any JB algebra which is the dual of a Banach space is called a *JBW algebra*[10]. Any JBW algebra has a unit denoted by $1\!\!1$. The system $\mathcal{J}(\mathcal{A})$ of all idempotent elements in a JBW algebra $\mathcal{A}$ forms a complete orthomodular[14,16] lattice, and $\mathcal{A}$ is the norm-closed linear hull of $\mathcal{E}$. For $E \in \mathcal{J}(\mathcal{A})$, $\{E\mathcal{A},E\}:=\{\{E,X,E\}|X \in \mathcal{A}\}$ is a subalgebra of $\mathcal{A}$ and is a JBW algebra with $E$ being the unit element. The idempotent elements in $\{E\mathcal{A},E\}$ coincide with the idempotent elements in $\mathcal{A}$ below $E$ and generate $\{E\mathcal{A},E\}$ as their norm-closed linear hull; $\{E,F,E\}$ lies in the norm-closed convex hull of $\{D \in \mathcal{E}|D \leq E\}$ as well as in $\mathcal{A}_+$, and $\{E,F,E\}=0$ if and only if $E \circ F=0$ $(E,F \in \mathcal{J}(\mathcal{A}))$.

A linear functional $\varphi : \mathcal{A} \to I\!\!R$ is called positive if $\varphi(X^2) \geq 0$ for all $X \in \mathcal{A}$. A positive linear functional $\varphi$ is bounded with $\|\varphi\| = \varphi(1\!\!1)$. For each $0 \neq X \in \mathcal{A}$ there exists a positive linear functional $\varphi$ with $\varphi(X) \neq 0$. The restriction of a positive linear functional $\varphi$ with $\varphi(1\!\!1)=1$ to $\mathcal{J}(\mathcal{A})$ provides a state on $\mathcal{J}(\mathcal{A})$.

**THEOREM 6.1:** *Let $\mathcal{A}$ be a JBW algebra. The system $\mathcal{J}(\mathcal{A})$ of idempotent elements in $\mathcal{A}$ is a UCP space if and only if $\mathcal{A}$ does not contain any type $I_2$ direct summand*[10].

Proof: If $\mathcal{A}$ contains a type $I_2$ direct summand, it follows from LEMMA 3.3 that $\mathcal{J}(\mathcal{A})$ is not a UCP space.

We now assume that $\mathcal{A}$ is a JBW algebra without any type $I_2$ direct summand. Then each state $\mu$ on $\mathcal{J}(\mathcal{A})$ has a unique extension $\hat{\mu}$ to a positive linear functional on $\mathcal{A}$[3,4,6,7,9,12,18,19]. This follows from results by Bunce and Wright[3,4] that finally provided a solution of the so-called Mackey problem for the JBW case, after it had been solved for the $W^*$-case by Christensen[6] and Yeadon[18,19], and after the early pioneering work by Gleason[9]. Note that these results do not hold in the type $I_2$ case.

Now let $\mu$ be any state on $\mathcal{J}(\mathcal{A})$ with $\mu(E)>0$ for some $E \in \mathcal{J}(\mathcal{A})$. We assume that a conditional probability $\mu_E$ exists. Let $\hat{\mu}$ and $\hat{\mu}_E$ be the extensions to positive linear functionals. Then for $F \in \mathcal{J}(\mathcal{A})$:

$$F = \left\{E,F,E\right\} + 2E' \circ (E \circ F) + E' \circ F, \quad \text{and} \quad \mu_E(F) = \hat{\mu}_E\left(\left\{E,F,E\right\}\right),$$

since $\hat{\mu}_E(E')=0$ and the Cauchy-Schwarz inequality imply that $\hat{\mu}_E(E' \circ (E \circ F))=0$ and $\hat{\mu}_E(E' \circ F)=0$. Now $\{E\mathcal{A},E\}$ is generated by the idempotents below $E$ and therefore

$$\mu_E(F) = \frac{1}{\mu(E)}\hat{\mu}\left(\left\{E,F,E\right\}\right). \qquad (*)$$

If a conditional probability $\mu_E$ exists, it must have this shape. So the uniqueness is proved. Since, on the other hand, $\{E,F,E\} \geq 0$ for idempotents $E,F$ in any JB-algebra, the above expression defines a conditional probability, and the existence of the conditional probability is proved as well.

The equation (∗) in the above proof gives us the shape of the conditional probabilities and implies that, in a JBW algebra $\mathcal{A}$, an event $F \in \mathcal{J}(\mathcal{A})$ is statistically predictable under an event $E \in \mathcal{J}(\mathcal{A})$ with $I\!\!P(F|E)=s$ if and only if the equation $\{E,F,E\}=sE$ holds in $\mathcal{A}$. We have $I\!\!P(F|E)=1$ if and only if $E \leq F$, and $I\!\!P(F|E)=0$ if and only if $E \leq F'$. This follows, since $\{E,F,E\}=E$ implies $E \leq F$. Furthermore,

$$I\!\!P\big(F\big|E_1,\ldots,E_n\big)=s$$

becomes equivalent to:

$$\Big\{E_1,\big\{E_2,\big\{\ldots,\big\{E_{n-1},\{E_n,F,E_n\},E_{n-1}\big\},\ldots\big\},E_2\big\},E_1\Big\}$$
$$= s\,\Big\{E_1,\big\{E_2,\big\{\ldots,\{E_{n-1},E_n,E_{n-1}\},\ldots\big\},E_2\big\},E_1\Big\}$$

We thus get purely algebraic expressions of the state-independent conditional probabilities. If $E$ and $F$ are atoms in a JBW algebra, then $I\!\!P(E|F)=I\!\!P(F|E)$, since $\{E,F,E\}=sE$ and $\{F,E,F\}=tF$ imply

$$s^2 E = \{E,F,E\}^2 = \{E,\{F,E,F\},E\} = t\{E,F,E\} = stE$$

and then $s=t$ or $s=0$. If $s=0$, $\{E,F,E\}=0$; hence $E \circ F=0$ and $\{F,E,F\}=0$, i.e. $t=0$.

**THEOREM 6.2:** *If $E$ and $F$ are compatible events in a JBW algebra without type $I_2$ part such that $E$ is statistically predictable under $F$, then either $F \leq E$ and $I\!\!P(E|F)=1$ or $F \leq E'$ and $I\!\!P(E|F)=0$.*

Proof: Using (∗), the compatibility implies: $E=\{F,E,F\}+\{F',E,F'\}$ and $F=\{E,F,E\}+\{E',F,E'\}$. From $\{F',E,F'\}=E-2E \circ F+\{F,E,F\}$ and $\{E',F,E'\}=F-2E \circ F+\{E,F,E\}$ we then get:

$$\{F,E,F\}=E \circ F=\{E,F,E\}$$

or, equivalently,

$$F \circ (E \circ F)=E \circ F=E \circ (E \circ F).$$

The statistical predictability means that $\{F,E,F\}=sF$ with $s=I\!\!P(E|F)$. Hence

$$E \circ (E \circ F)=E \circ F=\{F,E,F\}=sF$$

and

$$sE \circ F=E \circ (sF)=E \circ (E \circ F)=E \circ F,$$

which implies that either $s=1$ or $E \circ F=0$. In the latter case, we have $F \leq E'$ and $s=I\!\!P(E|F)=0$.

THEOREM 6.2 means that, in a JBW algebra, non-trivial cases of statistical predictability are possible only with events that are not compatible.

## 7. The quantum-mechanical Hilbert space model

Note that the self-adjoint part of any $W^*$-algebra[15] (von Neumann algebra) and the self-adjoint bounded linear operators on a complex Hilbert space as well as on a real or quaternionic Hilbert space form JBW algebras with the Jordan product $X \circ Y:=(XY+YX)/2$. In these cases: $\{X,Y,X\}=XYX$, and an event $E$ is statistically predictable under the event sequence $F_1,\ldots,F_n$ if and only if $F_1 F_2 \ldots F_n E F_n \ldots F_2 F_1 = s\, F_1 F_2 \ldots F_n \ldots F_2 F_1$ with a real number $s$. Then $I\!\!P(E|F_1,\ldots,F_n)=s$.

Thus, in these cases, DEFINITION 4.1 coincides with the concept of statistical predictability introduced in [13] where it was derived from an investigation of the Lüders - von Neumann quantum measurement process, but where the interpretation as state-independent conditional probability could not be provided.

However, on the one hand, the type $I_2$ cases must be excluded here, and on the other hand, there are exceptional[10] JBW algebras (not of type $I_2$) that do not have a representation on any Hilbert space and cannot be embedded in the self-adjoint part of any $W^*$-algebra.

With the standard model of quantum mechanics, the events are identified with the orthogonal projections on a complex Hilbert space $\mathscr{H}$, i.e. with the idempotent self-adjoint bounded linear operators on $\mathscr{H}$. They form a UCP space unless $dim\mathscr{H}=2$. We assume $dim\mathscr{H}>2$.

We already know that $I\!P(E|F)$ exists if $F$ is an atom. With the standard model of quantum mechanics, the atoms are the orthogonal projections on one-dimensional linear subspaces of the Hilbert space $\mathscr{H}$. Now let $F$ be an atom and $\xi$ be a vector in $F\mathscr{H}$ with $\|\xi\|=1$. Then

$$FEF = |\xi\rangle\langle\xi|E\xi\rangle\langle\xi| = \langle\xi|E\xi\rangle F \quad \text{and} \quad I\!P(E|F) = \langle\xi|E\xi\rangle.$$

If $E$ is a projection on a one-dimensional space as well and $\eta\in E\mathscr{H}$ with $\|\eta\|=1$, then

$$I\!P(E|F) = \langle\xi|\eta\rangle\langle\eta|\xi\rangle = |\langle\eta|\xi\rangle|^2.$$

Thus, we arrive at an expression familiar from quantum mechanics where the square of the absolute value of the inner product of two Hilbert space vectors carrying norm one is interpreted as a probability, but we are able to better understand the background of this statistical interpretation of quantum mechanics since knowing that the above expression is a state-independent conditional probability.

Non-trivial examples of the statistical predictability (i.e. $0<I\!P(E|F)<1$) can now easily be found by a proper selection of $\mathscr{H},\xi$ and $\eta$. These examples involve atoms. Can $I\!P(E|F)$ also exist and be different from 0,1 if $F$ is not an atom? It can, which is shown by the two matrices

$$E = \begin{pmatrix} 1 & 0 & 0 & 0 \\ 0 & 1 & 0 & 0 \\ 0 & 0 & 0 & 0 \\ 0 & 0 & 0 & 0 \end{pmatrix} \quad \text{and} \quad F = \frac{1}{2}\begin{pmatrix} 1 & 0 & 1 & 0 \\ 0 & 1 & 0 & 1 \\ 1 & 0 & 1 & 0 \\ 0 & 1 & 0 & 1 \end{pmatrix}$$

that represent orthogonal projections in the four-dimensional unitary space. Since $EFE = \frac{1}{2} E$, we get $I\!P(F|E) = \frac{1}{2}$.

Now let $E$ and $F$ be any two orthogonal projections on $\mathscr{H}$. Using (∗) and the identity $\{X,Y,X\}=XYX$, we get that $E$ and $F$ are compatible (DEFINITION 5.1) if and only if $E=FEF+F'EF'$ and $F=EFE+E'FE'$, which is equivalent to $EF=FE$, i.e. $E$ and $F$ commute. The probabilistically defined compatibility thus becomes identical with the algebraically defined concept of compatibility in the quantum-mechanical Hilbert space model.

## 8. Quantum interference

Let $D,E,F$ be three events in $\mathcal{J}(\mathcal{A})$, where $\mathcal{A}$ is the self-adjoint part of a W*-algebra without type $I_2$ direct summand. Let $D$ be an atom, $\rho_D$ the unique state with $\rho_D(D)=1$, i.e. $\rho_D(G)=I\!P(G|D)$, and $\hat{\rho}_D$ the unique linear extension of $\rho_D$ to a positive linear functional on $\mathcal{A}$. Assume that $E\neq0$ and $I\!P(E|D)\neq0$. Then:

$$I\!P(F|D,E)=\rho_D(F/E)=\hat{\rho}_D(EFE)/\hat{\rho}_D(E).$$

We are now interested in the case where $E$ is the sum of two orthogonal non-zero events $E_1$ and $E_2$:

$$I\!P\big(F|D,E_1+E_2\big)=\frac{1}{I\!P(E_1+E_2|D)}\big(\hat{\rho}_D\big(E_1FE_1\big)+\hat{\rho}_D\big(E_2FE_2\big)+\hat{\rho}_D\big(E_1FE_2\big)+\hat{\rho}_D\big(E_2FE_1\big)\big)$$

$$=I\!P\big(F|D,E_1\big)\frac{I\!P\big(E_1|D\big)}{I\!P\big(E_1+E_2|D\big)}+I\!P\big(F|D,E_2\big)\frac{I\!P\big(E_2|D\big)}{I\!P\big(E_1+E_2|D\big)}+2\frac{\mathrm{Re}\,\hat{\rho}_D\big(E_1FE_2\big)}{I\!P\big(E_1+E_2|D\big)}$$

If neither $E_1$ nor $E_2$ is compatible with $F$ (i.e. commutes with $F$), the last term on the right-hand side of this equation need not vanish (although $E_1\perp E_2$) and, moreover, can be negative as well as positive. This term is responsible for a certain oscillating deviation from the sum of the first two terms on the right-hand side of the equation (this sum represents the classical case), i.e. the non-Boolean probabilities include wave-like interference phenomena as observed with quantum-physical particles.

For instance, consider the 2-slit experiment with a quantum-physical particle. Let $D$ be the event that the particle owns a certain fixed linear momentum, $E_k$ ($k=1,2$) the events that the particle passes through slit 1 and 2, respectively, and $F$ the event that the particle is detected at a certain fixed location $x$ behind the screen with the two slits. Then these probabilities own the following interpretations:

| | |
|---|---|
| $I\!P(F|D,E_1)$ | is the probability that the particle will be detected at $x$ if slit 1 is open and slit 2 is closed. |
| $I\!P(F|D,E_2)$ | is the probability that the particle will be detected at $x$ if slit 1 is closed and slit 2 is open. |
| $I\!P(F|D,E_1+E_2)$ | is the probability that the particle will be detected at $x$ if both slits are open. |
| $I\!P(E_1|D)$ | is the probability that the particle flies through slit 1. |
| $I\!P(E_2|D)$ | is the probability that the particle flies through slit 2. |
| $I\!P(E_1+E_2|D)$ | is the probability that the particle flies through either slit 1 or slit 2 (which is identical with the probability that it does not hit the screen). |

We now assume that $D,E_1,E_2,$ and $F$ are orthogonal projections on 1-dimensional subspaces spanned by the normalized vectors $\xi,\eta_1,\eta_2$ and $\psi$, respectively, in a complex Hilbert space where $\eta_1$ and $\eta_2$ are orthogonal. Then

$$I\!P(F|D,E_1+E_2)=\frac{\big|\langle\xi|\eta_1\rangle\langle\eta_1|\psi\rangle+\langle\xi|\eta_2\rangle\langle\eta_2|\psi\rangle\big|^2}{\big|\langle\xi|\eta_1\rangle\big|^2+\big|\langle\xi|\eta_2\rangle\big|^2}.$$

The right-hand side of this equation exactly reproduces the quantum-mechanical superposition of state vectors, which is the standard explanation for interference phenomena.

## 9. Quantum measurement

That observations of nature behave as strange as described by the UCP space model in sec. 4 may appear unbelievable to common sense, but yet such observations have been the daily business of quantum physicists for eighty years now. For instance, let $\vec{x}_o, \vec{x}_1,...,\vec{x}_n$ be space axes and let $\lambda_o,\lambda_1,...,\lambda_n$ be possible values of the spin of a quantum physical particle measured along theses axes. Let $E$ denote the particle property that its spin along axis $\vec{x}_o$ is $\lambda_o$, and let $F_k$ denote the particle property that its spin along axis $\vec{x}_k$ is $\lambda_k$. Measurements of these particle properties then exactly reproduce the behavior described in sec. 4. The probability $I\!P(E|F_1,...,F_n)$ depends on the angles among the axes $\vec{x}_o,\vec{x}_1,...,\vec{x}_n$ and on $\lambda_o,\lambda_1,...,\lambda_n$. By a proper selection of $\vec{x}_o,\vec{x}_1,...,\vec{x}_n$ and $\lambda_o,\lambda_1,...,\lambda_n$, any value in the unit interval can be achieved for this probability.

A quantum measurement can thus be understood as a mere observation of a certain property of a physical system. Outcomes of quantum measurements are events, and $I\!P(E|F_1,...,F_n)$ is the probability of the outcome $E$ with a future measurement testing $E$ versus $E'$, after a series of measurements ($k=1,...,n$) testing $F_k$ versus $F_k'$ has given the results $F_k$. This probability depends on the results of the measurement series only, it does not depend on any initial state of the physical system under consideration. If $E$ is not statistically predictable under $F_1,...,F_n$, the knowledge of the outcomes of the measurement series is not sufficient for any prediction concerning the occurrence of $E$ or $E'$ with a future measurement.

Quantum measurement phenomena that appear strange to common sense can thus perfectly be explained by using non-Boolean probabilities, where it is not necessary to take into account the measuring apparatus. However, this does not at all mean that everything becomes quite simple now since non-Boolean probabilities will remain difficult to understand with common sense and may, as quantum mechanics itself, result in a conflict with the assumption that an objective absolute physical reality exists.

The UCP space model does not use the wave-functions or Hilbert space vectors $\psi$ which therefore do not own any interpretation within this model and the role of which reduces to a mathematical auxiliary tool available in special cases only. This may decrease the physical and philosophical significance of the so-called collapse of the wave-function with a quantum measurement. Many misunderstandings of quantum theory result from the assumption that $\psi$ represents an actual state of the system under observation. With the UCP space model, however, $\psi$ itself does not have any meaning, but it is the projection on the linear span of $\psi$ which represents the observer's information on the system that was achieved as the outcome of a measurement. Considering EPR[8] experiments, this may remove the alleged contradiction between Einstein's locality principle and quantum mechanics, but not the conflict between quantum-mechanics and *"realism"*.

## 10. Conclusions

The UCP space model presented here is a non-Boolean extension of the classical probability model. It reproduces typical quantum phenomena, but is easier to interpret than the quantum-mechanical Hilbert space formalism, and it exhibits a particular phenomenon, the non-trivial state-independent conditional probabilities or statistical predictability, which may provide opportunities for a new understanding of quantum measurement.

The problems with the assumption that there is an objective absolute physical reality behind quantum measurement now become problems of a non-Boolean extension of probability theory, or of a non-Boolean logic since the state-independent conditional probabilities depend only on the underlying logico-algebraic structure of the events/propositions and may therefore be regarded more a logical than a stochastic phenomenon. These probabilities themselves have an objective character and thus differ from classical probabilities the origin of which always lies in the observer's subjective lack of information.

We have encountered two cases where the quantum-mechanical Hilbert-space model and the UCP space model do not overlap each other. The first one does not cover the so-called exceptional Jordan algebras. The second one does not include the 2-dimensional Hilbert spaces or type $I_2$ algebras which play a major role in quantum mechanics since describing the single quantum bit as well as the spin $\hbar/2$. However, spin $\hbar/2$ particles always own further properties (e.g. linear momentum, mass, charge, etc.); the combined consideration of all particle properties requires dimensions higher than 2 and is then covered by the UCP model.